\documentclass[lettersize,journal]{IEEEtran}
\usepackage{amsmath,amsfonts}
\usepackage{algorithmic}
\usepackage{algorithm}
\usepackage{array}
\usepackage[caption=false,font=normalsize,labelfont=sf,textfont=sf]{subfig}
\usepackage{textcomp}
\usepackage{stfloats}
\usepackage{url}
\usepackage{verbatim}
\usepackage{graphicx}
\usepackage{xcolor}
\usepackage{cite}
\usepackage{ulem}

\begin{document}

\title{Spaceflight KID Readout Electronics for PRIMA}

\author{Thomas Essinger-Hileman$^1$, %~\IEEEmembership{Staff,~IEEE,}
C. Matt Bradford$^2$, Patrick Brown$^1$, Sean Bryan$^3$, Jesse Coldsmith$^1$, Jennifer Corekin$^1$, Sumit Dahal$^1$, Thomas Devlin$^1$, Marc Foote$^2$, Draisy Friedman$^1$, Alessandro Geist$^1$, Jason Glenn$^1$, Christopher Green$^1$, Tracee Jamison-Hooks$^3$, Kevin Horgan$^1$, Jared Lucey$^1$, Philip Mauskopf$^{3,4}$, Lynn Miles$^1$, Sanetra Bailey Newman$^1$, Gerard Quilligan$^1$, Cody Roberson$^3$, Adrian Sinclair$^1$, Salman Sheikh$^1$, Eric Weeks$^3$, Christopher Wilson$^1$, Travis Wise$^1$
        % <-this % stops a space
\thanks{$^1$NASA Goddard Space Flight Center, Greenbelt, MD 20771 USA}
\thanks{$^2$NASA Jet Propulsion Laboratory, Pasadena, CA 91109 USA}
\thanks{$^3$School of Earth and Space Exploration, Arizona State University, Tempe, AZ 85287 USA}
\thanks{$^4$Department of Physics, Arizona State University, Tempe, AZ 85281 USA}% <-this % stops a space
\thanks{Manuscript received TBD; revised TBD.}}

\markboth{\textit{IEEE} Transactions on Applied Superconductivity}%
{Shell \MakeLowercase{\textit{et al.}}: A Sample Article Using IEEEtran.cls for IEEE Journals}

%\IEEEpubid{0000--0000/00\$00.00~\copyright~2025 IEEE}

\maketitle

\begin{abstract}
We present the design and testing of a prototype multiplexing kinetic inductance detector (KID) readout electronics for the PRobe far-Infrared Mission for Astrophysics (PRIMA) space mission. PRIMA is a Probe-class astrophysics mission concept that will answer fundamental questions about the formation of planetary systems, the co-evolution of stars and supermassive black holes in galaxies, and the rise of heavy elements and dust over cosmic time. The readout electronics for PRIMA must be compatible with operation at Earth-Sun L2 and capable of multiplexing more than 1000 detectors over 2.5 GHz bandwidth. The electronics must also be capable of switching between the two instruments, which have different readout bands: the hyperspectral imager (PRIMAger, 2.6-4.9 GHz) and the spectrometer (FIRESS, 0.4-2.4 GHz). The PRIMA readout electronics use high-heritage SpaceCube digital electronics with a build-to-print SpaceCube Mini v3.0 board using a radiation-tolerant Kintex KU060\footnote{Neither the U.S. Government nor NASA endorse or recommend any commercial products, processes, or services. Reference to or appearance of any specific commercial products, processes, or services by trade name, trademark, manufacturer, or otherwise, does not constitute or imply its endorsement, recommendation, or favoring by the U.S. Government or NASA. The views and opinions of the authors do not necessarily state or reflect those of the U.S. Government or NASA, and they may not be used for advertising or product endorsement purposes.} field programmable gate array (FPGA) and a custom high-speed digitizer board, along with RF electronics that provide filtering and power conditioning. We present the driving requirements for the system, as well as the hardware, firmware, software, and system-level design that meets those requirements.
\end{abstract}


\section{Introduction}
\label{sect:intro}  % \label{} allows reference to this section
The rapid development in the past two decades of kinetic inductance detectors (KIDs) for the far-infrared (far-IR) with sensitivities approaching the background limit from zodiacal light for moderate-resolution ($R \sim 100$) spectroscopy (noise equivalent power,
$\mathrm{NEP} \sim 10^{-19}~\text{W}\sqrt{\text{Hz}}$) promises to open new windows on the obscured universe \cite{day2024_25umKIDs, Baselmans_2022}. Large arrays of KIDs operating from a space platform, enabled by their inherent ability to be frequency multiplexed, would improve upon the sensitivity of previous far-IR observatories (\textit{e.g.}, \cite{neugebauer1984iras, kessler1996iso, werner2004spitzer, murakami2007akari, pilbratt2010herschel, gehrz2011sofia}) by orders of magnitude and would fill a critical gap in current astronomical capabilities between the James Webb Space Telescope (JWST)\cite{gardner2023jwst} at wavelengths $\lambda < 28~\text{\textmu m}$ and the Atacama Large Millimeter/Submillimeter Array (ALMA) at wavelengths $\lambda > 300~\text{\textmu m}$.

The PRobe Mission for Astrophysics (PRIMA)~\cite{prima_overview_glenn_jatis} is a far-IR space telescope proposed as a NASA Astrophysics Probe Explorer. PRIMA has a 1.8-meter primary mirror maintained at 4.5~K that couples to two instruments: the Far-Infrared Enhanced Survey Spectrometer (FIRESS)\cite{firess_bradford_jatis}, which has a high-resolution mode when coupled to a Fourier Transform Module (FTM), and a hyperspectral and polarimetric imager (PRIMAger).\cite{primager_ciesla_jatis}. The design and fabrication of KIDs for PRIMA demonstrate the sensitivity required for the space environment\cite{foote2024_PRIMAKIDS, haileydunsheath2025_charcPRIMAKIDS}. Here we describe warm spaceflight readout electronics for PRIMA that build on hardware and firmware approaches from ground-based and balloon-borne instruments.\cite{sinclair2020_blast, masi19olimpo}

\section{Requirements Flow Down}
To maintain the background-limited far-IR sensitivity of the instrument we require the readout electronics noise to be sub-dominant when compared to the intrinsic photon and detector noise. We can estimate noise specifications from the detector properties; the noise equivalent power (NEP), the fractional frequency responsivity ($dx/dP$ that converts power, P, to fractional frequency shift in the resonators, $x=df/f$), and the resonator quality factors, where $Q_r = Q_i Q_c /(Q_i + Q_c)$ is the total quality factor which depends on the internal ($Q_i$) and coupling ($Q_c$) quality factors. The noise spectral density (NSD) in dBc/Hz is estimated from these detector properties as, 
\begin{equation}
    \text{NSD} = 20\text{log}_{10}(\text{NEP}\frac{dx}{dP}\frac{2Q_r^2}{Q_c}).
\end{equation}

Which can be derived by taking the ratio of the detector output noise power with the bias tone power \cite{Sipola2019,Mauskopf_2018}. Typical values of these parameters for PRIMA detectors \cite{haileydunsheath2025_charcPRIMAKIDS,foote2024_PRIMAKIDS} are $\mathrm{NEP} \sim 10^{-19}~\text{W}\sqrt{\text{Hz}}$, $dx/dP \sim 2.7 \times 10^{10}\ 1/\text{W}$, $Q_r=10^4$, $Q_c=1.1\times 10^4$, which yield intrinsic detector $\mathrm{NSD} \sim -86$ dBc/Hz.
Note that this is an approximate form for when the bias tone is exactly on resonance and for other offsets the full derivatives of the resonator forward transmission should be used. We require that the NSD of the readout electronics be sub-dominant to the detector NSD. This can be estimated from the following equation \cite{WP509Xilinx2019,Gordon_2016_Readout,Rantwijk2016},
\begin{equation}
\begin{split}
    \text{NSD} = -6.02\cdot\text{ENOB} - 1.76 - 10\text{log}_{10}(f_s/2) \\
    + 10\text{log}_{10}(N_{tones}) + CF + \Delta P_\text{bias}
\end{split}
\label{eqn:NSD}
\end{equation}
where ENOB is the effective number of bits, $f_s$ is the sampling rate, $N_{tones}$ is the number of simultaneously generated tones, $CF$ is the waveform crest factor, and $\Delta P_\text{bias}$ the tone bias power variation. We require each readout channel to be capable of simultaneously reading out 1008 detectors (the maximum in a PRIMA readout chain) along with a small number of blind tones spread across 2.5~GHz of instantaneous bandwidth. The crest factor for a sum of 1008 tones with random phases and frequencies is around 12 dB, however impedance mismatches along the analog path can increase this by a few dB. These three terms are constant in the equation above leaving the design variables to be the ENOB and $\Delta P_\text{bias}$. The ENOB is determined by the high-speed digitizer choices, clocking performance, and digital signal processing gains. Due to the fact that each KID requires slightly different bias powers this variation is an added bit of dynamic range required by the readout electronics. In addition to the optimal bias power variations, the radio-frequency (RF) chain flatness from the output of the digital-to-analog converter (DAC) down to the detectors and back to the analog-to-digital converter (ADC) will contribute to the $\Delta P_\text{bias}$ term. 

Given estimated levels for both RF system flatness and bias power variations, we set a requirement on digital system noise in the absence of the $\Delta P_\text{bias}$ term of $< -95$~dBc/Hz, which is achievable with available high-speed spaceflight ADC and DAC chips (See Section~\ref{sec:flight_arch} for details). In addition to the white noise level, PRIMA aims for the readout electronics to be stable in time with $1/f$ noise from slow drifts that are sub-dominant to the white noise for frequencies in the time streams above 5~Hz after removal of common mode drifts shared by all detectors. Common mode removal processing will occur post downlink on the ground on modern CPUs.

\begin{figure*}
\begin{center}
\begin{tabular}{c}
\includegraphics[width=0.9\textwidth, clip=true, trim=0.95in 2.1in 0.1in 1.3in]{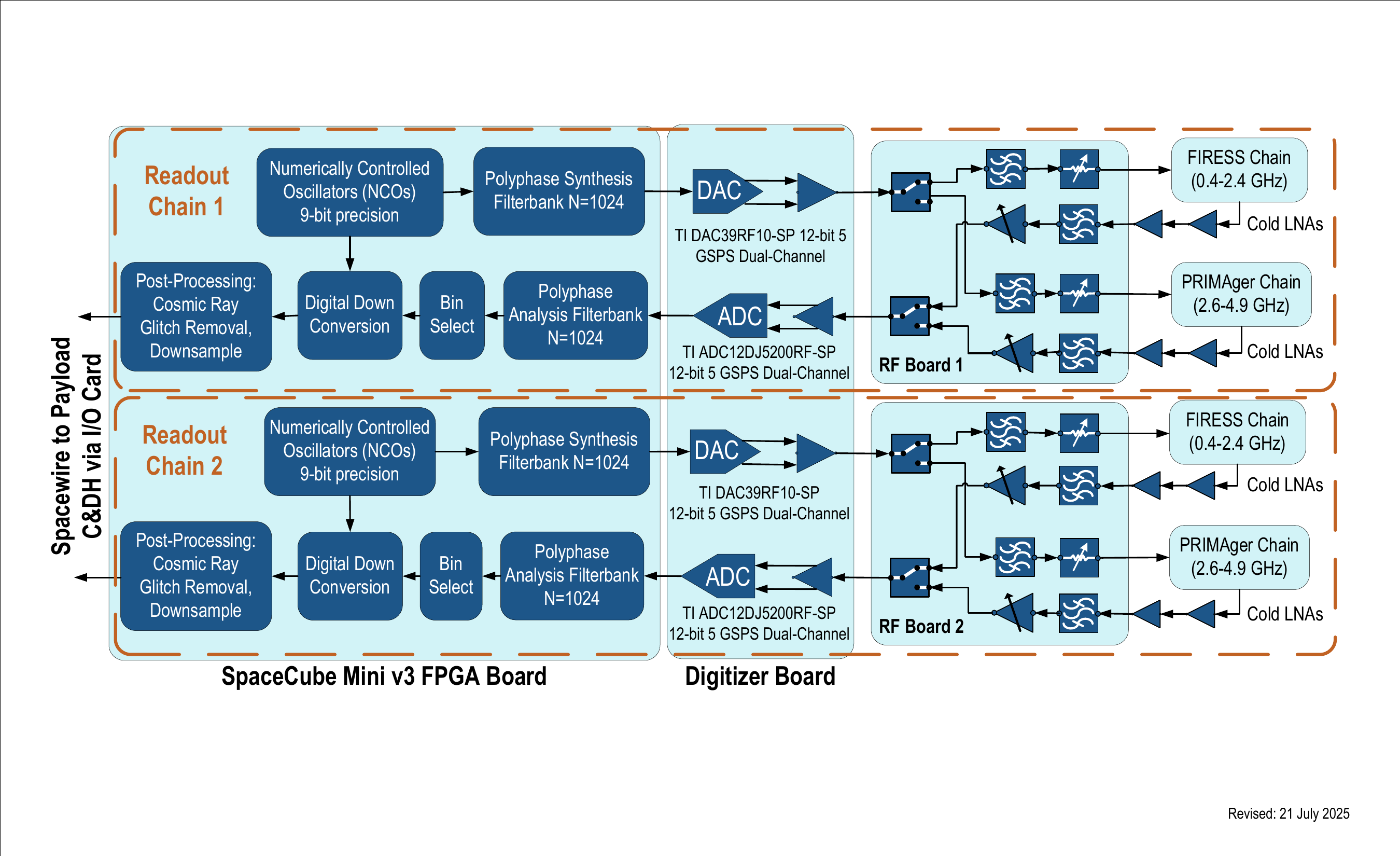}
\end{tabular}
\end{center}
\caption 
{ \label{fig:readout_diagram}
Block diagram of the PRIMA KID readout chain. The 8 chains are switched between FIRESS and PRIMAger with filtering to allow readout of FIRESS in the first Nyquist zone and PRIMAger in the second Nyquist zone of the ADC and DAC when sampling at 5 Gsps. A high-heritage SpaceCube Mini v3.0 FPGA board houses a Xilinx Kintex Ultrascale KU060 FPGA and has high-speed interfaces with the ADC/DAC digitizer board over a backplane. The FPGA and digitizer boards can each read out 2 detector chains as illustrated. The radio-frequency (RF) electronics provide switching between the two instruments, filtering, and adjustment of overall gain for system noise optimization. Digitized detector timestreams are sent over SpaceWire to the spacecraft control and data handling (C\&DH) computer via an input-output (I/O) card.} 
\end{figure*} 

\section{PRIMA Flight System Architecture}
\label{sec:flight_arch}
Frequency multiplexing readout electronics for the KID arrays are shared by FIRESS and PRIMAger via RF switching with one instrument operating at a time. In each of our 8 signal chains, a comb of tones, one for each KID in the array, is transmitted from a synthesis polyphase filterbank (PFB) on a Xilinx Kintex KU060 Ultrascale FPGA. The $N=1024$ length synthesis PFB provides coarse channelization of the outgoing tones, while a set of 32 numerically controlled oscillators (NCOs) provide a set of time-variable sinusoids to the PFB bins for fine channelization. The current architecture supports up to 32 tones to be generated within the same bin. Tone placement and recovery precision of better than 10~kHz is required at the low-frequency end of the FIRESS (0.4-2.4~GHz) readout band, set by the KID resonator quality factors of $\sim 10^4$. The NCOs with 16 bit phase accumulators feed the 1024 point synthesis PFB providing a frequency resolution of 5~GHz$/1024/2^{16-1} \approx 149$~Hz.
%Fine channelization of the PFB bins, with width of 5~GHz$/1024=4.88$~MHz, by $2^9$ provides tone placement precision of 9.54~kHz.

The comb is converted to analog using a DAC\footnote{Texas Instruments, Inc., DAC39RF10-SP, https://www.ti.com/product/DAC39RF10-SP}, filtered to reject out-of-band images, attenuated, and sent through coaxial cables to the KID arrays. After interacting with the KID array, the modified comb returns through coaxial cables with amplification by low-noise amplifiers (LNAs) at 18~K, 100~K, and ambient temperature before being digitized by an ADC.\footnote{Texas Instruments, Inc., ADC12DJ5200RF-SP, https://www.ti.com/product/ADC12DJ5200RF} An analysis PFB for coarse channelization followed by a digital down conversion fine channelization measures the in-phase (I) and quadrature (Q) components of the received tones with the same frequency resolution as the synthesis PFB. The resultant I and Q time-ordered-data (TOD) are accumulated at the output of the digital down converters and then sent to a fractional rate resampling filter to achieve the desired science data rate ($\sim$100-700~Hz), depending on the observing mode and needs of individual science observations. 

\begin{figure*}
\begin{center}
\begin{tabular}{c}
\includegraphics[width=0.9\textwidth, clip=true, trim=0in 2in 0in 1in]{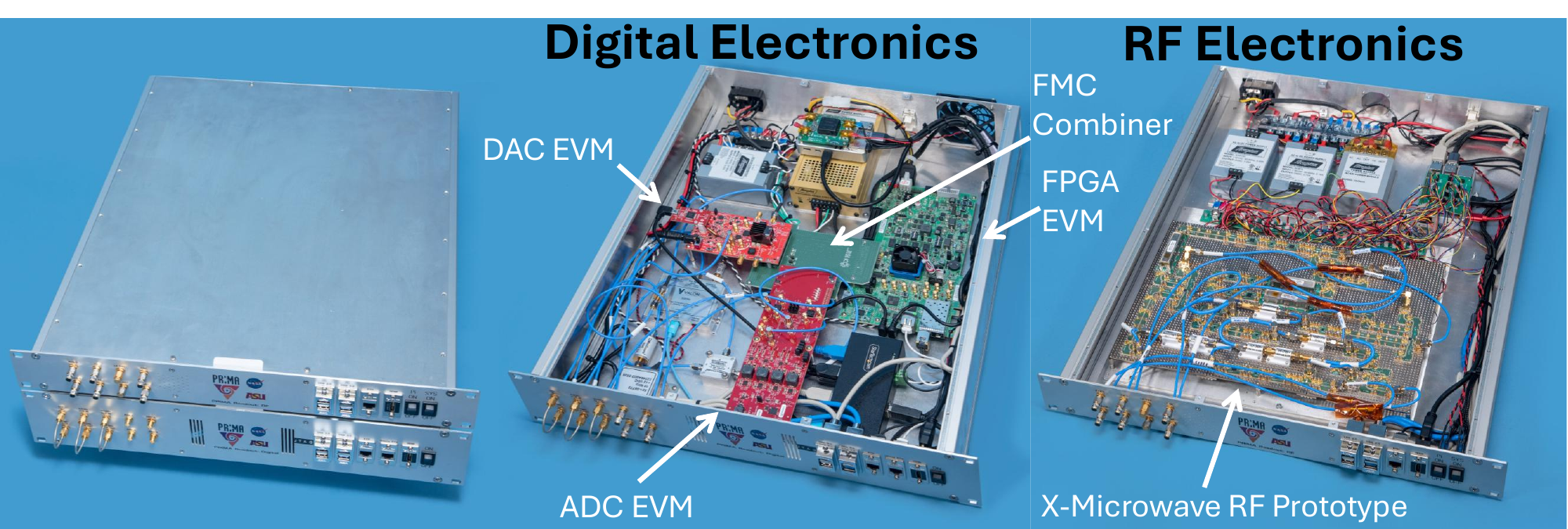}
\end{tabular}
\end{center}
\caption 
{ \label{fig:prototype_photo}
Photographs of the PRIMA prototype electronics. (\textit{Left panel:}) Combined rack-mount readout electronics with radio-frequency (RF) electronics on top and digital electronics on the bottom. (\textit{Middle panel:}) Digital electronics prototype with lid opened with ADC, DAC, and FPGA evaluation modules (EVMs) labeled, along with the FPGA Mezzanine Card (FMC) combiner board that connects them together. The enclosure also contains a reference clock, Raspberry Pi for control, USB hub, and power supplies. (\textit{Right panel:}) The RF electronics box with lid opened and the Quantic X-Microwave prototype circuit labeled. The RF enclosure houses a Raspberry Pi for control and power supplies.} 
\end{figure*} 

The ADC and DAC rates of 5 giga-samples per second (Gsps) place the FIRESS readout band of 0.4–2.4 GHz in the first Nyquist zone and the PRIMAger 2.5–5.0 GHz readout band in the second Nyquist zone. An RF board provides switching between FIRESS and PRIMAger, with filtering for band selection and variable attenuation to scale the outgoing and incoming tones. The band limiting filters within the RF electronics board were chosen such that the in-band slope and ripple are less than 1 dB and out of band aliases will be attenuated by at least 30 dB. Modular digital readout electronics use high-heritage SpaceCube boards.~\cite{Petrick15_SpaceCube, Wilson15_SpaceCube, Brewer20_SpaceCube, Sabogal17_SpaceCube, Perryman21_SpaceCube, Kanekal2019_SpaceCube_MERiT_CeREs} The SpaceCube Mini v3.0 FPGA board and associated power supply cards are build-to-print and currently flying on the STP-H9-SCENIC mission on the ISS~\cite{Geist23_SpaceCube}. PRIMA requires a custom ADC/DAC board, input-output (I/O) card, backplane, and enclosure, which are adaptations of existing designs.~\cite{Dubayah20_GEDI, Sun20_GEDI}

Fig.~\ref{fig:readout_diagram} shows the signal flow in a single readout chain. The algorithms have been used for over a decade in ground-based and balloon-borne instruments.~\cite{Gordon_2016_Readout, Sinclair2022_CCAT_Readout} One unique aspect of the PRIMA readout electronics is the need to perform cosmic-ray glitch removal on-board before sending to the ground. Available daily downlink volumes for the spacecraft cannot support TOD from the full 8064 detectors in the FIRESS instrument at the maximum output sampling rate (10~kHz). The PRIMA KIDs have time constants of approximately 1~ms, which is the characteristic time of cosmic ray glitches, making it advantageous to flag and remove glitches in the 10~kHz data before further accumulation and resampling. Initial tests of deglitching algorithms on simulated and laboratory data show that this is feasible with matched filters and threshold detection similar to those used in photon-counting applications in the visible through the X-ray. Work is on-going to integrate cosmic ray deglitching into the PRIMA firmware.

\section{Phase A Prototype}
PRIMA is currently completing a one-year Phase A study to mature the design concept in anticipation of an Astrophysics Probe selection in 2026. During Phase A, prototype readout electronics were designed and built, which demonstrate the full signal processing chain of the flight design in rack-mount enclosures (See Fig.~\ref{fig:prototype_photo}). The analog RF electronics occupy one enclosure and implement the flight design envisioned for the RF Boards (See Fig.~\ref{fig:readout_diagram}) using Quantic X-Microwave X-MWblocks.\footnote{Quantic X-Microwave, Inc., https://quanticxmw.com/}

The digital electronics occupy a second enclosure, which contains evaluation modules (EVMs) for the candidate PRIMA flight ADC, DAC, and FPGA chips. A custom FPGA Mezzanine Card (FMC) combiner board was designed and fabricated to connect the three EVMs. For the phase A prototype a Raspberry Pi\footnote{Raspberry Pi 5, https://www.raspberrypi.com/products/raspberry-pi-5/} provides command and data handling functions, in flight a radiation tolerant C$\&$DH computer will be used. The system also contains a 10 MHz oven-controlled crystal oscillator reference clock to reduce system drift.\footnote{501-08772, https://www.quanticwenzel.com/} 

\begin{figure}
\begin{center}
\begin{tabular}{c}
\includegraphics[width=0.45\textwidth, clip=true, trim=0in 0in 0in 0in]{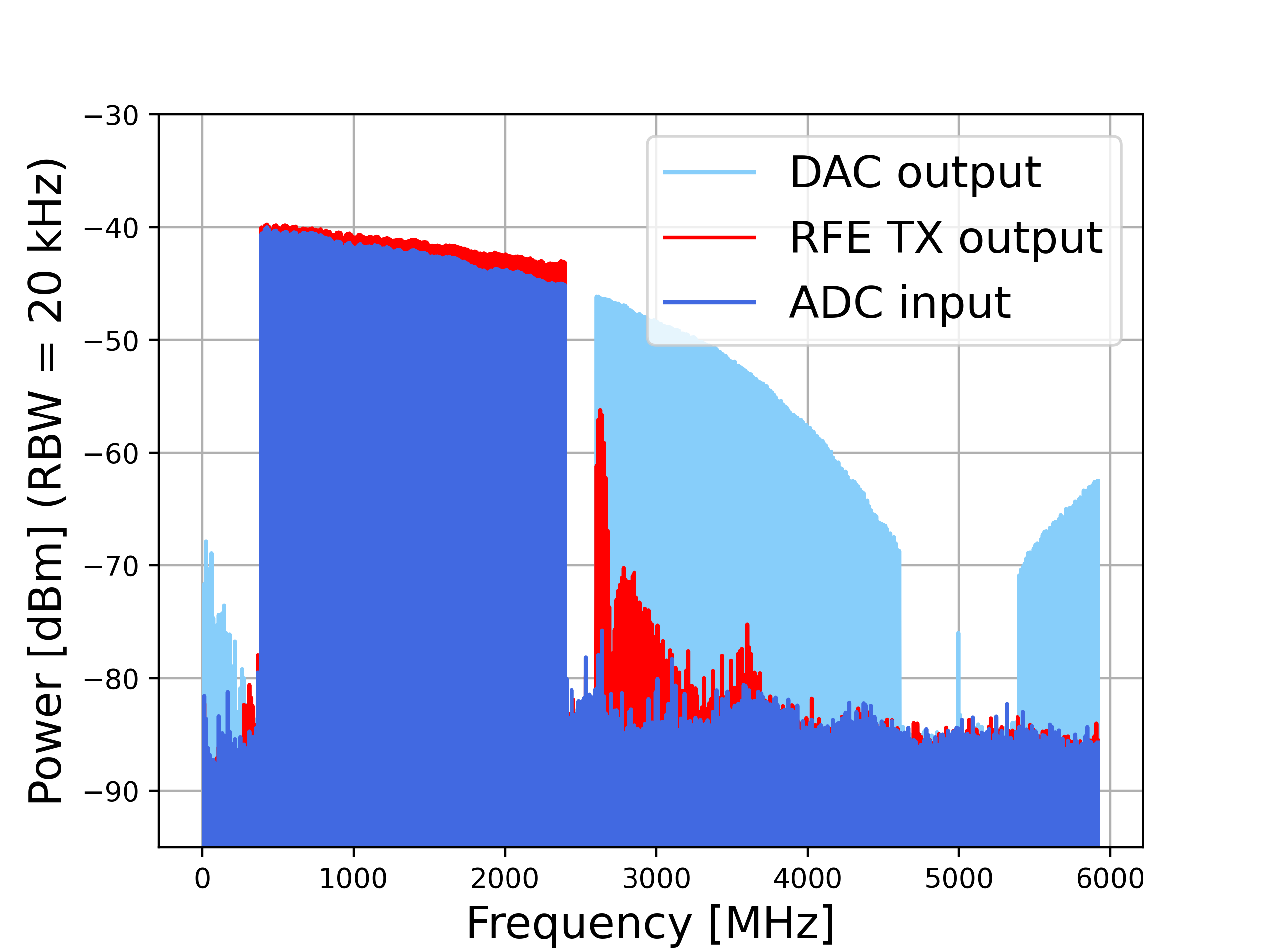}
\end{tabular}
\end{center}
\caption 
{ \label{fig:tone_output}
Frequency comb generated by the phase A prototype digital and RF electronics for the FIRESS band (0.4 to 2.4 GHz) measured in three different configurations. The first of which shown in light blue is measured directly from the DAC port on the digital electronics front panel and shows the first, second, and start of the third Nyquist images. Shown in red is the same comb after it has passed through the RF electronics prototypes transmit chain while in "FIRESS science" mode. The third configuration in dark blue was measured at the input to the ADC. This represents an RF electronics loopback configuration with a 20 dB attenuator between the TX and RX ports as a stand in for cryostat loss. 
} 
\end{figure}

Here we present initial performance of the assembled systems, including demonstration of tone output across the FIRESS band (0.4-2.4~GHz) with the expected bandpass. Figure~\ref{fig:tone_output} shows a comb of narrow tones spaced equally across the FIRESS band at three points in the system: (1) output directly from the DAC and showing the expected frequency response of the DAC, including images of the tones in the second and third Nyquist zones, (2) output from the RF Electronics after analog filtering, and (3) at the input to the ADC after additional filtering. 

\begin{figure}
\begin{center}
\begin{tabular}{c}
\includegraphics[width=0.45\textwidth, clip=true, trim=0in 0in 0in 0in]{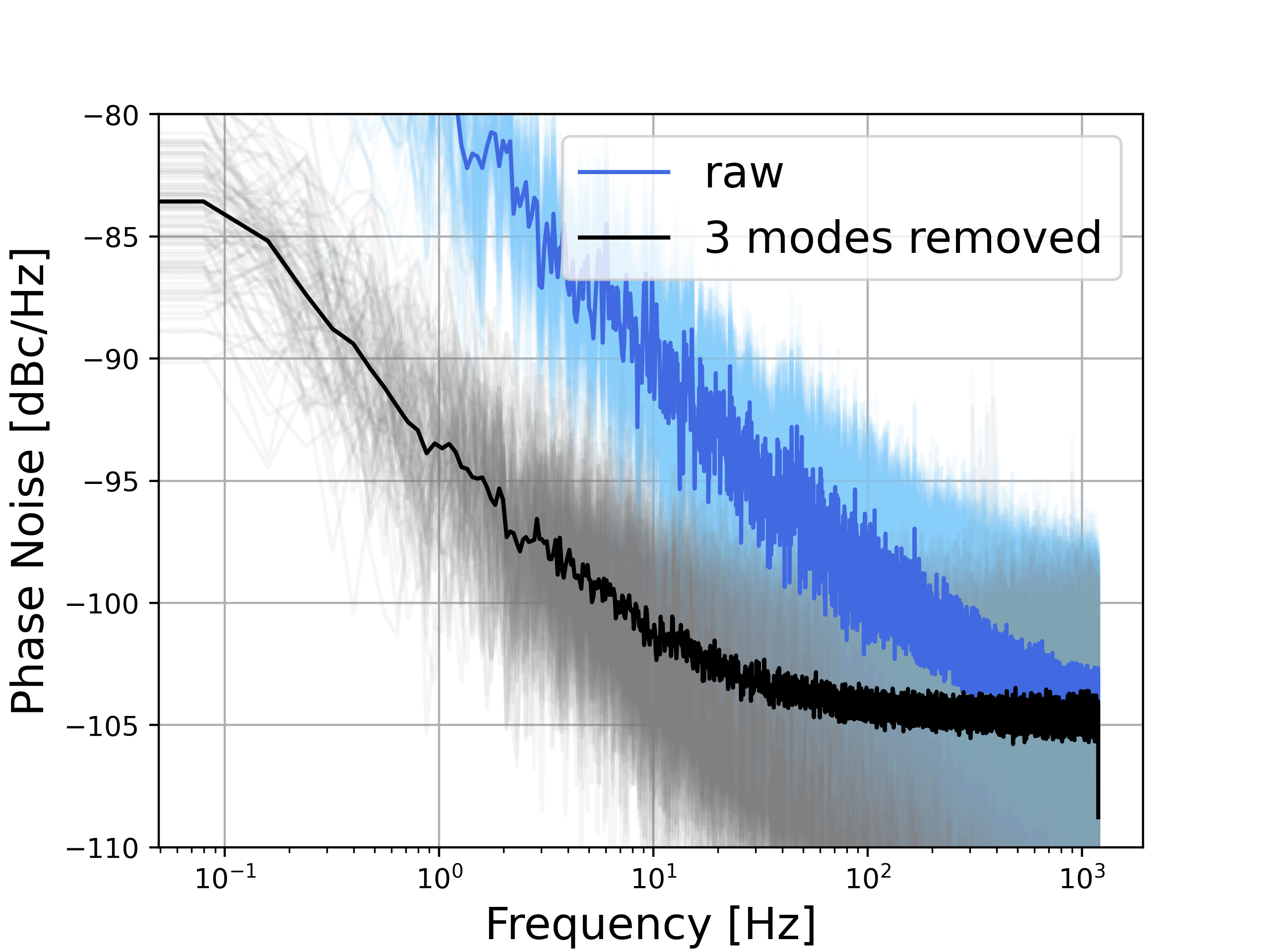}
\end{tabular}
\end{center}
\caption 
{ \label{fig:psd} Noise spectral density (NSD) of 100 sets of time-ordered data (TOD) taken with the PRIMA prototype readout, corresponding to tones spread across the FIRESS 0.4-2.4~GHz band. The blue curve is the NSD of the raw data, while the black curve shows the noise after removal of 3 common modes using a principal component analysis. The bold curve in each case is the median in each frequency bin of the set of faint 100 per-tone NSDs shown. This demonstrates a white noise level of $-105$~dBc/Hz and $1/f$ noise of approximately $-95$~dBc/Hz at a frequency of 1~Hz.} 
\end{figure}

In-phase and quadrature components (or alternatively amplitude and phase) of the TOD are taken at the output of the digital receiver while in a loopback configuration. The NSD of the TOD for 100 tones is shown in Figure ~\ref{fig:psd}. Raw data show high $1/f$ noise, which is significantly mitigated by removal of as few as three common modes using a principal component analysis (PCA). This demonstrates that the system achieves a white noise level $\sim -105$~dBc/Hz and $1/f$ noise of approximately $-95$~dBc/Hz at a frequency of 1~Hz for 100 tones. The expected noise level for the 1008 tones needed to address a FIRESS array is approximately 10~dB higher (see Eq.~\ref{eqn:NSD}).

The phase A prototype readout electronics will be tested with PRIMA KID arrays in dilution refrigerators at both Goddard Space Flight Center and the Jet Propulsion Laboratory. The cryogenic RF chain at Goddard has been set up in a flight like configuration such that the low noise amplifiers intended to read out each RF chain are thermally controlled to sit at 18 and 100 K, corresponding to the available thermal stages on the spacecraft. The results of these tests with detectors in a flight like cryogenic configuration with the prototype readout electronics will be the subject of a future publication. 

\section{Conclusion}
Warm readout electronics for the proposed PRIMA far-IR Astrophysics Probe Explorer mission concept have been designed to enable addressing up to 1008 detectors per readout chain along with a set of blind diagnostic tones. Switching between the two instruments (FIRESS and PRIMAger) on board the spacecraft is done in analog RF electronics that also provide filtering and power conditioning for optimal detector biasing. The readout band for the two instruments is different, with FIRESS being read out 0.4--2.4~GHz and some PRIMAger chains being read out over 2.6--4.9~GHz. These correspond to the first and second Nyquist zones of the 5~Gsps direct-sampling digital readout electronics. A rack-mount prototype has been built from commercial parts with space-grade counterparts, with the RF electronics composed of Quantic X-Microwave X-MWblocks and the digital electronics of evaluation modules of flight candidate ADC, DAC, and FPGA chips. Initial testing of the assembled prototype electronics demonstrate the ability to output 100 tones over the FIRESS band with a loopback white noise floor of -105 dBc/Hz (-95 dBc/Hz when extrapolated to the required 1008 detectors in a FIRESS readout chain) and $1/f$ noise of $-95$~dBc/Hz at 2 Hz.

\section*{Acknowledgments}
The authors gratefully acknowledge useful discussions with numerous PRIMA colleagues that shaped this work. We also thank Logan Foote for sharing principle component analysis software. Researchers at the Jet Propulsion Laboratory, California Institute of Technology, are under a contract with the National Aeronautics and Space Administration (80NM0018D0004)

\bibliography{references}
\bibliographystyle{IEEEtran}

\end{document}